\documentstyle[prd,eqsecnum,aps]{revtex}
\begin{document}
\title{Improved initial data for black hole collisions}
\author{Carlos O. Lousto\thanks{Electronic Address:
lousto@mail.physics.utah.edu} and Richard H. Price}
\address{
Department of Physics, University of Utah, Salt Lake City, UT 84112}
\maketitle
\begin{abstract}
Numerical relativity codes now being developed will evolve initial
data representing colliding black holes at a relatively late stage
in the collision. The choice of initial data used for code
development has been made on the basis of mathematical
definitiveness and usefulness for computational implementation. By
using the ``particle limit'' (the limit of an extreme ratio of
masses of colliding holes) we recently showed that the standard
choice is not a good representation of astrophysically generated
initial data.  Here we show that, for the particle limit, there is a
very simple alternative choice that appears to give excellent
results.  That choice, ``convective'' initial data is, roughly
speaking, equivalent to the start of a time sequence of
parameterized solutions of the Hamiltonian constraint; for a
particle in circular orbit, it is the initial data of the steady
state solution on any hypersurface. The implementation of related
schemes for equal mass holes is discussed.
\end{abstract}

\section{Introduction}

Recently the numerical computation of gravitational waves from
coalescing black holes has received considerable attention and effort
for closely related reasons. (i) The problem is a testbed for numerical
relativity and has been the center of a collaboration of eight
universities\cite{samreview}. (ii) The results are necessary to
understand details, and perhaps even gross features of coalescing
black holes, presently viewed by many as the most promising source of
gravitational waves detectable by earth based
instruments\cite{schutzbet,fh}. (iii) The problem is interesting in
its own regard as the first solution of Einstein's equations for
strong field interactions without simplifying symmetry.

The job of the numerical codes under development will be to evolve
forward in time from an initial value solution on a starting
hypersurface.  There are many ways that initial data can be chosen to
represent the starting state of two black holes.  If the black
holes are far apart on the initial hypersurface, then any reasonable
choice of hypersurface data will suffice. Unfortunately the
instabilities in evolution codes mean that the codes can only run for
a relatively short time, and initial data will have to be specified on
a hypersurface as late as possible.

For the most part, the initial data used to the present has been that
put forth by York and Bowen, and
coworkers\cite{yorkpiran,yorkjmp,bowenyork}.  This scheme starts by
requiring that the initial geometry be conformally flat and that the
extrinsic curvature have a property called ``longitudinal,'' equivalent
to the condition that it be derivable from a vector potential. The
solution is then made inversion symmetric, so that topological throats
representing the black holes connect two isometric asymptotically flat
universes. (This step of making the hypersurface data inversion
symmetric turns out to destroy ``longitudinality''\cite{rauber}.  )

The advantages of this approach to initial data include the following:
(i) Since it is widely adopted in the numerical relativity community,
comparison of results of different groups is facilitated. (ii) It is
specific and definitive and the numerical methods of solving for these
initial data are highly developed\cite{cook}. (iii) The inversion
symmetry gave a simple inner boundary condition for evolution codes.
(iv) An argument could be made that the initial data was free of, or
had a minimal content of, radiation encoded (in some sense) in the 
initial data themselves, so that the outgoing radiation found in 
a spacetime evolution should represent only the radiation due to the 
dynamical interaction of the black holes.

The Bowen-York (hereafter BY) scheme served very well in the role
intended for it. One should ask, however, whether it is the
appropriate scheme for computations supporting the search for
gravitational waves. There are strong reasons for concluding that the
answer is ``no.'' One immediate reason is that the Kerr geometry
itself cannot be put in this scheme; a constant time slice of the Kerr
geometry is not conformally flat. Even a single spinning BY hole will
therefore radiate gravitational waves as it settles down into
its  Kerr final state\cite{snglspn}.

A second rather direct reason comes from recent calculations we
carried out for radiation from a particle falling radially into a
Schwarzschild hole\cite{lp1,lp2}, hereafter Paper I and Paper II.  In
these calculations, the ``particle'' was taken as a perturbation of
the background black hole spacetime and computations could be done
relatively easily for a variety of choices of data on an initial
hypersurface. Furthermore, evolved data could be extracted on later
hypersurfaces and compared with various prescriptions (BY etc.) for
hypersurface data. In the approach used, based on the perturbation
formalism of Moncrief\cite{moncrief}, the initial and evolved data was
represented by a spacetime function $\psi$ that is totally gauge
invariant, so that the interpretation of the physics was not tainted
by questions of coordinate choices.

The calculations were used to study a set of hypersurface data that
somewhat generalize the BY data. Those sets considered were all
conformally flat and longitudinal (CFL). In practice this gave, for
each multipole of the perturbation, a complete set of differential
equations that had to be satisfied by the hypersurface geometry and
extrinsic curvature (or, in practice, $\psi$ and its time derivative
$\dot{\psi}$). The only remaining freedom is the choice of boundary
conditions (consistent with asymptotic flatness). These choices are
equivalent to the choice of conditions deep inside the throats
representing the holes, i.e., in the asymptotically flat universes at
the end of the throats away from ``our'' asymptotically flat universe.
We focused on two choices of these boundary conditions. Once choice
was the choice of the BY data itself. The second was the choice that
the solution at the horizon be ``frozen'' or ``matched'' to its
initial value. This approach was based on the fact that very near the
background horizon a simple analytic solution could be given for the
evolution of the data on the initial hypersurface to subsequent
hypersurfaces.  The solution for $\psi$ was simply that it was
``frozen'' to retain its initial value at the horizon. The solution
for $\dot{\psi}$ reflected more detail, and was said to be ``matched''
to the initial solution at the horizon. We could also choose to have
$\psi$ ``frozen,'' but we could choose the boundary conditions for
$\dot{\psi}$ to be the BY conditions, and so forth.

For a particle falling from rest at a large initial distance, we found
that the solution that evolved, and the CFL data on later
hypersurfaces were in serious disagreement. If evolved data were
replaced by CFL data on a late hypersurface the resulting waveforms
and radiated energies were drastically altered.  We found furthermore:
(i) In the class of CFL hypersurface data the best choice is the
``frozen-matched'' choice, but that the difference between this and
other CFL choices is small compared to the disagreement with the
evolved solution. (ii) The disagreement between evolved hypersurface
data and CFL data does not lie, in any obvious way, in the ``radiation
content'' of the data, but rather in the details of the data near the
location of the particle. (iii) The disagreement between the true and
CFL data was much more severe in $\dot{\psi}$ than in $\psi$.

In the present paper we look again at the problem of head on collision
of two nonspinning holes in the particle limit, and ask whether there
is a more successful alternative prescription for hypersurface data.
More specifically, let us consider a configuration of black holes at
late times that evolved from a precursor configuration at large
separation, and let us suppose that we know the positions and momenta
of the holes on the late time hypersurface. Is there a prescription
for the data on the late hypersurface which is in good agreement with
the ``true'' hypersurface data (i.e., the data evolved from the
large-separation precursor), and which gives an accurate prediction of
the radiated energy and waveform?
The answer to this question seems to be that there is, and the
prescription, at least in the particle limit, is a very simple one:
that the data be ``convective,'' in a sense to be defined below.

In Sec.\ IIA we sketch the mathematics of the Moncrief formalism we
use and of our notation and conventions, and we present some results
of evolved data vs.\ CFL data; this sketch will be quite brief; a
reader interested in the details will find them in Paper II.  The
proposal for convective data is discussed in general in Sec.\ IIB, and
is implemented in a particular manner, for radial infall, in Sec.\
IIC. Numerical results from this implementation are given in Sec.\
IID. A discussion of the implications and possible extensions of this
work is presented in Sec.\ III.

\section{Astrophysical initial data}
\subsection{Review of particle results for BY initial data}
\subsubsection{Moncrief-Zerilli formalism}

The numerical results we will present below will be for a particle
infalling radially into a Schwarzschild black hole.
Due to the axial symmetry of our problem, and the absence of rotation,
the perturbations are pure even parity.  To describe the
perturbations, we use the Regge-Wheeler\cite{rw} notation, but not the
Regge-Wheeler gauge.  In this notation the even parity metric
perturbations are decomposed into scalar, vector and tensor spherical
harmonics.  with coefficient functions $ H_0^\ell, H_1^\ell, H_2^\ell,
h_0^\ell, h_1^\ell, K^\ell, G^\ell$\,, depending on $r,t$ the standard
radial and time coordinates of the Schwarzschild background. For
simplicity, we drop the multipole index $\ell$ on these perturbation
functions.

The perturbation coefficients 
$H_2,
h_1,K,G$ describe only the perturbed 3-geometry, and are independent of 
the choice of shift and lapse. In terms of these, Moncrief\cite{moncrief}
defines the totally gauge invariant function 
\begin{equation}\label{psidef}
  \psi(r,t)=\frac{r}{{\lambda}+1}\left[
    K+\frac{r-2M}{{\lambda}r+3M}\left\{ H_2-r\partial K/\partial r
    \right\} \right]
+\frac{2}{r}(r-2M)\left(r^2\partial G/\partial r-2h_1
\right)\ ,
\end{equation}
where we have used Zerilli's normalization for $\psi$ and his notation
\begin{equation}{\lambda}\equiv(\ell+2)(\ell-1)/2\ .
\end{equation}

The basic wave equation for an infalling particle is given in Paper I
as 
\begin{equation}\label{rtzerilli}
-\frac{\partial^2\psi}{\partial t^2}
+\frac{\partial^2\psi}{\partial r*^2}-V_{\ell}(r)\psi=
{\cal S}_\ell (r,t)\ .
\end{equation}
Here $r^*\equiv r+2M\ln(r/2M-1)$ is the Regge-Wheeler\cite{rw}
``tortoise'' coordinate and $V_\ell$ is the Zerilli potential 
\begin{equation}\label{zpotential}
V_\ell(r)=\left(1-\frac{2M}{r}\right)
\frac{2\lambda^2(\lambda+1)r^3+6\lambda^2Mr^2+18\lambda M^2r+18M^3
}{r^3(\lambda r+3M)^2 }\ .
\end{equation}
The right hand side of Eq.\ (\ref{rtzerilli})
is a source term constructed from the particle stress energy\cite{lp1,lp2}.
For a point particle of proper mass $m_0$, the
stress energy is given by
\begin{equation}\label{tmunu}
T^{\mu\nu}=(m_0/U^0)U^\mu U^\nu \delta[r-r_p(t)]\delta^2[\Omega]/r^2\ ,
\end{equation}
where $U^\mu$ is the particle 4-velocity, and $r_p(t)$ gives the
radial position of the particle as a function of coordinate time,
starting from $r_p=r_0$ at $t=0.$

The total energy radiated in the $\ell$th multipole, after a hypersurface
at $t_0$ is 
\begin{equation}
{\rm Energy}_\ell=
\frac{1
}{64\pi
}
\frac{(\ell+2)!
}{(\ell-2)!
}\ \int_{t_0}^\infty\left(\dot{\psi}^\ell\right)^2dt\ .
\end{equation}

\subsubsection{Functional freedom in the hypersurface data}

An even parity multipole perturbation of the 3-geometry has the four
functional (of $r,t$) degrees of freedom contained in $h_1,H_2,K,G$,
but these must satisfy the (perturbed, multipole) Hamiltonian
constraint, a single second order differential equation containing the
particle source.  The choice of a conformally flat 3-geometry reduces
the problem to a single functional degree of freedom (since, in an
appropriate coordinate system, $h_1=G=0$ and $H_2=K$) satisfying the
single Hamiltonian constraint.  The only freedom is the choice of the
constants specifying the homogeneous solution of the Hamiltonian
constraint.  One of those constants is fixed by the condition that the
solution be well behaved at spatial infinity.  The remaining constant
can be specified to ``freeze'' the horizon value of $\psi$ at its
initial value, to make the solution inversion symmetric, or in other
ways.  If the requirement of conformal flatness is dropped, then any
sufficiently smooth function can be added to a particular solution
$\psi$ that solves the Hamiltonian constraint.

For $\dot{\psi}$ things are a bit more complicated. For an even parity
perturbative extrinsic curvature there are four functions of $r,t$,
analogous to the four functions $h_1, H_2, K, G$ specifying the
3-geometry perturbations. There are two second order equations,
equivalent to $G_{tr} =8\pi T_{tr}$ and $G_{t\theta} =0$.  If we
require that the extrinsic curvature be longitudinal, then the
extrinsic curvature must be derivable from a vector potential
$\vec{W}$, and for a given even parity multipole the functional
degrees of freedom are $W^r(t,r)$ and $W^\theta(t,r)$. Since these
functions must satisfy two second order differential equations, we
have no remaining functional freedom.  We have only to choose the four
constants that specify how much our solution contains of the four
homogeneous solutions of the momentum constraints. Two of these
constants are fixed by the requirement that the extrinsic geometry be
well behaved at spatial infinity. The remaining two constants can be
chosen to make the solution of the BY type, to make the solution
``match'' the initial solution at the horizon, etc.  If the
requirement of longitudinality is dropped, then any sufficiently
smooth function can be added to any particular solution $\dot\psi$ of the
momentum constraints.

\subsubsection{Results for CFL hypersurface data}

Here we briefly review the numerical results found for the CFL
prescription. In the results presented here a particle started its
infall from Schwarzschild radial coordinate $r_p=r_0=15(2M)$, where
$M$ is the mass of the spacetime (in the usual $c=G=1$ units).  At
subsequent times, when the radial location $r_p$ of the particle is
smaller, and the time label of a $t=$\, constant hypersurface is
larger, we can extract the evolved $\psi$, replace it with a prescribed
CFL $\psi$ etc. 

Figure \ref{fig:reven} shows the result that is most important. At
various times (parameterized by the particle location $r_p^*$ at these
times) the numerical evolution was stopped, and the evolved quadrupole
$\psi$ and $\dot{\psi}$ were replaced by the CFL prescribed data
representing the particle position and momentum on this hypersurface.
Evolution forward from this hypersurface is then done numerically. The
resulting energy is then numerically computed. The true $\ell=2$
energy for infall from $r_0=15(2M)$ is $0.0164(m_0^2/2M)$. If we evolve
the prescribed data from a later surface we see that we get reasonably
good agreement with the true energy for $r_p^*$ larger than around
$5(2M)$, corresponding to $r_p$ around $4(2M)$. For smaller $r_p$ the
use of prescribed CFL data badly overestimates the radiated energy.

Comparisons of the true (i.e., evolved) quadrupolar $\psi,\dot{\psi}$
and the prescribed quadrupolar $\psi,\dot{\psi}$ are shown in the next
two figures. Both are for a particle starting from rest at initial
separation $r_0=15(2M)$ at $t=0$. For both, the comparison is made at
$t=96.9(2M)$.  This corresponds to the particle being at $r_p=2(2M)$,
and having momentum $P=1.27m_0$. A comparison of $\psi$ is given in
Fig.\ \ref{fig:revpsi} in which the solid curve shows the evolved
$\psi$ and the dashed curve is the CF $\psi$ for a particle of mass
$m_0$, at location $r_p=2(2M)$, with momentum $P=1.27m_0$.
The corresponding comparison for $\dot{\psi}$ is shown in Fig.\ 
\ref{fig:revpsidot}. From these comparisons it appears that the CF
solutions give a reasonably good approximation to $\psi$ near the
particle, but in the strong field region of the background, the CFL
form of $\dot{\psi}$ has a very different shape near the particle than
that of the evolved $\dot{\psi}$.  We note here that the evolved
$\dot\psi$ will, in general, have a $\delta[r-r_p]$ behavior at the
location of the particle.  This behavior is exactly included in both
the CFL and the convective (see below) prescriptions.

\subsection{Proposed astrophysical initial data}

Suppose that our particle source starts at some position
$r_0,\theta_0,\varphi_0$. 
On some later hypersurface, let us call it the $t=t_\Sigma$ hypersurface,
suppose that we know the coordinate position $x^j_p$ and the momentum
$P^j$ for the particle.  (Here latin superscripts, $j,k,\ldots$ are
spatial indices running from 1 to 3.)  From the geodesic motion of the
particle we can find $x^j_p$ and $P^j$ at any time $t$.  Let us now
take $\Psi(x^k;x^k_p)$ to be a form of $\psi$ that is ``appropriate''
to the position (and momentum) of the particle.  This could be, for
example, the $\psi$ corresponding to a CF 3-geometry for the particle.
The next step is to promote this solution to a function of time, by
putting in the explicit time dependence $x^k_p[t]$.
(Note that the resulting time-dependent function $\Psi(x^k;x^k_p[t])$ is
not based on a solution of Einstein's equations, and will not solve
the Zerilli equation.)
Our proposal is to take the hypersurface data to be:
\begin{equation}\label{psiprop}
\psi|_{t_\Sigma}=\Psi(x^k;x^k_p[t_\Sigma]) 
\end{equation}\begin{equation}\label{psidotprop}
\dot{\psi}|_{t_\Sigma}=\left.\frac{\partial\Psi(x^k;x^k_p[t])}
{\partial t}\right|_{t_\Sigma}
=\left[\frac{dx^j_p[t]}{dt}\ 
\frac{\partial\Psi(x^k;x^k_p[t])}{\partial x^j_p}\right]_{t_\Sigma}\ .
\end{equation}
With this choice we are, in effect, saying that $\psi$ is changing in
time only in order to adjust to the new particle position and
momentum. The solution, in a sense, is dragged or transported to a new
position. We call this the ``convective'' choice of $\dot{\psi}$.

Initial data, of course, must be chosen so that it solves the
constraint equations. But here (so far) we have been discussing the
choice only of the single gauge invariant function $\psi$ on the
hypersurface, and its time derivative.  Since the constraints leave
a functional degree of freedom in the initial 3-geometry and extrinsic
curvature, $\psi$ and $\dot{\psi}$ can be freely specified.

One motivation for this choice of hypersurface data is that the
disagreement between evolved and prescribed initial data seems to be
rooted in the CFL $\dot{\psi}$ much more than in $\psi$. The proposed
convective choice of hypersurface data allows us to maintain the CF
form of $\psi$, but find a new form of $\dot{\psi}$.
A second motivation has to do with motion in which the hypersurface
data is ``obvious,'' a particle in circular orbit at frequency
$\Omega$ around a hole. In this case the appropriate approximation
(neglecting radiation reaction to lowest order) is for the spacetime
geometry to be periodic, with period $2\pi/\Omega$. More specifically,
in terms of the coordinates $t,r,\theta,\varphi$ of the Schwarzschild
(or Kerr) background, all spacetime functions should have the form
$f(t,r,\theta,\varphi)=F(r,\theta,\varphi-\Omega t)$.

If we denote the position of the particle by $r_p=r_0, \theta_p=\pi/2,
\varphi_p=\varphi_0+\Omega t$. The solution for $\psi$ must therefore have
the functional form as
\begin{equation}
\psi(t,r,\theta,\varphi)=\Phi(r,\theta, \varphi-\Omega
t)=\Phi(r,\theta, \varphi-\varphi_p[t]+\varphi_0)\ .
\end{equation}
We can therefore take
\begin{equation}\label{Psioft}
\Psi(x^k;x^k[t])
=\Phi(r,\theta, \varphi-\varphi_p[t]+\varphi_0)\ .
\end{equation}
This solution clearly obeys Eq.\ (\ref{psidotprop}) on any
hypersurface $t_\Sigma$, since the only time dependence in $\Psi$ comes
through its dependence on $\varphi_p[t]$.
If on some hypersurface $t=t_\Sigma$,  one chooses initial data
\begin{equation}
\psi|_{t_\Sigma}=\Phi|_{t_\Sigma}\ ,\;\;\;
\dot{\psi}|_{t_\Sigma}=\left.-\Omega
\frac{\partial\Phi}{\partial\varphi}\right|_{t_\Sigma}\ ,
\end{equation}
then, in principle,
numerical evolution would give a purely periodic solution. For $x^k,P^k$
corresponding to circular motion, one could, on the
other hand, give other hypersurface data, such as CFL.
The result would be a burst of
initial radiation as the spacetime ``settles down''
to the periodic solution. If the circular orbit is meant to represent
a late stage in a binary coalescence which is adiabatically decaying due
to radiation reaction, then the periodic solution is obviously the
appropriate one, and an initial burst of radiation is an anomaly due to
the wrong choice of initial conditions.

\subsection{Convective  conformally flat data}

We now apply the idea of convective $\dot{\psi}$ to a particle of mass
$m_0$ falling radially into a
Schwarzschild hole.  As the appropriate form of $\psi$ on a
hypersurface we choose a conformally flat solution, with ``frozen''
horizon conditions. That is, on any hypersurface, the
$r^*\rightarrow-\infty$ limit of $\psi$ is the same as on the $t=0$
hypersurface. 

The meaning of this solution can be nicely described with the notation
developed in Paper II. For even parity $\ell$-pole perturbations of the
3-geometry, there are only two gauge invariant functions of $r$.  One
is the Moncrief $\psi$ defined by Eq.\ (\ref{psidef}). The second is
$I_{\rm conf}$, a gauge invariant measure of deviation of the
3-geometry from conformal flatness. (In the Regge-Wheeler\cite{rw}
gauge $I_{\rm conf}\equiv H_2-K$.)  For all CFL choices $I_{\rm conf}$ is
zero on the hypersurface. This choice exhausts the functional
freedom on the hypersurface, and the 3-geometry is then fixed by the
Hamiltonian constraint. In practice this means that $\psi$ satisfies a
second-order differential equation in the $r$ variable; the frozen
horizon condition fixes the solution to that equation.

We are then choosing the same hypersurface $\psi$ as a CFL choice. What is
crucial is the difference in our choice of $\dot{\psi}$. Here our
convective choice is tantamount to taking $\Psi[(x^k;x^k_p(t)]$ to be
conformally flat for all $t$. This corresponds to a form of $\psi$,
on each hypersurface, with $I_{\rm conf}=0$. It follows that 
\begin{equation}
\dot{I}_{\rm conf}|_{t_\Sigma}=0\ .
\end{equation}
In order to understand better the meaning of this condition it is
useful to relate it to the Hamiltonian constraint.  To derive this
relationship conveniently we momentarily specialize to the gauge
$h_0=h_1=G=0$ of Regge and Wheeler\cite{rw}. In this gauge the
Hamiltonian constraint is given by Zerilli\cite{zerilli2} as his
Eq. (C7a). For a particle at radial position $r_p$ this takes the form
\begin{eqnarray}\label{Gtt}
\left(1-\frac{2M}{r}\right) \frac{\partial^2K}{\partial r^2}
+\left(3-\frac{5M}{r}\right) \frac{1}{r}\frac{\partial K}{\partial r}
- \frac{\lambda}{r^2}K
- \frac{\lambda+2}{r^2}H_2-\left(1-\frac{2M}{r}\right)
\frac{1}{r}\frac{\partial H_2}{\partial r}=
\nonumber\\
-\kappa\ U^0\left(1-\frac{2M}{r}
\right)\frac{1}{r^2}\delta[r-r_p]\ ,
\end{eqnarray}
where $\kappa=2m_0\sqrt{4\pi(2\ell+1)}$, 
$m_0$ is the mass of the particle, and $U^0$ is its time
component of 4-velocity.

{}From Eq.\ (\ref{Gtt}) and its radial derivative, and from Eq.\
(\ref{psidef}) we can write
\begin{eqnarray}\label{energia}
I_{\rm conf}=(r-2M)\psi''+\frac{[(\lambda-3)r+9M]M}{(\lambda r+3M)r}\psi'
-\frac{[27M^3+24\lambda M^2r+3\lambda(3\lambda+1)Mr^2+2\lambda^2
(\lambda+1)r^3]}{(\lambda r+3M)^2r^2}\psi\nonumber\\
+\frac{\kappa\ U^0(1-2M/r)
[\lambda(\lambda+1)r^2-3M^2]}{(\lambda+1)(\lambda r+3M)^2}\delta[r-r_p]
-\frac{\kappa\ U^0(r-2M)^2}{(\lambda+1)(\lambda r+3M)}
\delta'[r-r_p]\ .
\end{eqnarray}
This result was derived in the Regge-Wheeler gauge as an intermediate step,
but is gauge-invariant.
In a similar way, from the momentum constraints (see Eqs.\ (C7b), (C7d) of
Zerilli\cite{zerilli2}) and their $r$-derivatives, we obtain
\begin{eqnarray}\label{momentum}
\dot I_{\rm conf}=(r-2M)\dot\psi''+\frac{[(\lambda-3)r+9M]M}{(\lambda r+3M)r}\dot\psi'
-\frac{[27M^3+24\lambda M^2r+3\lambda(3\lambda+1)Mr^2+2\lambda^2
(\lambda+1)r^3]}{(\lambda r+3M)^2r^2}\dot\psi\nonumber\\
+\frac{2\kappa\ U^0\dot r_p
[9M^2+2M(\lambda-3)r-\lambda(\lambda+3)r^2]}
{(\lambda+1)r^2(\lambda r+3M)^2}\delta[r-r_p]\\
+\frac{\kappa\ U^0\dot r_p(r-2M)
[9M^2+2M\lambda r-\lambda(\lambda+1)r^2]}
{(\lambda+1)r(\lambda r+3M)^2}\delta'[r-r_p]
+\frac{\kappa\ U^0\dot r_p(r-2M)^2}
{(\lambda+1)(\lambda r+3M)}\delta''[r-r_p]\ .\nonumber
\end{eqnarray}
Note that this equation can be obtained as the time derivative of Eq.\
 (\ref{energia}), but here we wish to emphasize its connection with
 the momentum constraints. The fact that there is consistency of the
 time derivative of the Hamiltonian constraint, and the spatial
 derivatives of the momentum constraints, is simply an expression of
 the fact that $G^{\ \nu}_{0\ \ ;\nu}=0$.

We can view Eq.\ (\ref{energia}) as telling us that $I_{\rm conf}$
vanishes, since the right hand side vanishes for conformally flat data
that satisfies the Hamiltonian constraint. Alternatively we can note
that in Eq.\ (\ref{energia}) if we put $I_{\rm conf}$ to zero, what
results is a second order radial differential equation for $\psi$. In
this sense, the condition $I_{\rm conf}=0$ and the Hamiltonian
constraint gives us (aside from boundary conditions) the initial value
of $\psi$.  In the same manner we can note that $\dot I_{\rm conf}=0$
in Eq.\ (\ref{momentum}) gives us a radial differential equation for
$\dot\psi$. From this point of view the condition $\dot I_{\rm
conf}=0$ and the momentum constraints are fixing the form of
$\dot\psi$. By contrast, in the CFL prescription, the form of
$\dot\psi$ is fixed by the momentum constraints and the condition of
longitudinality, roughly the condition that the conformal part of the
extrinsic curvature can be derived from a vector potential.

To see how our proposal works, let us first consider the radial infall of a
particle into a Schwarzschild hole. We have seen
in Paper II that there is a preferred choice, $\psi_f$, that
keeps  its value at the event horizon ``frozen''
\begin{eqnarray}\label{psi0}
\psi_f=\frac{2m(r_p)}{\lambda+1}\frac{
\sqrt{4\pi/(2\ell+1)}}{\lambda r+3M}r\sqrt{r/\bar{r}}
\times\Biggl[
\left(\lambda+1+{M\over r}+\sqrt{1-2M/r}\left(\ell+\sqrt{\bar{r}/r}\right)
\right)\Gamma_{hom}(r_p){(M/2)^{2\ell+1}\over\bar{r}_p^{\ell+1}\bar{r}^\ell}
+\nonumber\\
\left\{\begin{array}{c}
\left(\lambda+1+M/r+\sqrt{1-2M/r}\left(\ell+\sqrt{\bar{r}/r}\right)
\right)(\bar{r}_p/\bar{r})^\ell\\
\left(\lambda+1+M/r+\sqrt{1-2M/r}\left(\sqrt{\bar{r}/r}-\ell-1\right)
\right)(\bar{r}/\bar{r}_p)^{\ell+1}
\end{array}
\right\}\Biggr]\ ,
\end{eqnarray}
where
\begin{equation}
\bar{r}\equiv\left(\sqrt{r}+\sqrt{r-2M}\right)^2/4\ ,
\end{equation}
and where the upper expressions apply in the case $\bar r>\bar r_p$,
and the lower for $\bar r<\bar r_p$. Here, as in Paper II
\begin{equation}\label{uno}
\Gamma_{hom}(\bar{r}_p)={m(r_0)\over m(r_p)}\left({\bar{r}_p\over
\bar{r}_0}\right)^{\ell+1}\left(1+\Gamma_{hom}(r_0)\right)-1~,
\end{equation}
where $\Gamma_{hom}(r_0)$ represents the arbitrariness in choosing
the data on the initial hypersurface $t=0$ (when the particle started
at rest and was located at $r_p=r_0$), and 
\begin{equation}\label{finalm2m0}
m(r_p)={\textstyle\frac{1}{2}}m_0
\left(1+\frac{1}{\sqrt{1-2M/r_p}}\right)\epsilon_0\ .
\end{equation}
where $\epsilon_0=\sqrt{1-2M/r_0}$ for a particle starting at rest
when $r=r_0$, and $\epsilon_0=1/\sqrt{1-v^2_\infty}$ for a particle with
velocity $v_\infty$ at infinity, and $U^0=\epsilon_0/(1-2M/r_p)$.

{}From this expressions we compute the ``convective''
$\dot\psi_c=\dot r_p\partial_{r_p}\psi_f$
\begin{eqnarray}\label{psi0p}
&&\dot\psi_c=\frac{2m(r_p)\sqrt{4\pi/(2\ell+1)}}{\lambda+1}
\frac{r\sqrt{r/\bar{r}}}{\lambda r+3M}
\frac{\dot r_p}{(r_p-2M)}
\times\nonumber\\
&&\Biggl[
\left(\lambda+1+{M\over r}+\sqrt{1-2M/r}\left(\ell+\sqrt{\bar{r}/r}\right)
\right)\left((\ell+1)\sqrt{1-2M/r_p}+M/(r_p+r_p\sqrt{1-2M/r_p})\right)
{(M/2)^{2\ell+1}\over\bar{r}_p^{\ell+1}\bar{r}^\ell}
+\\
&&\left\{\begin{array}{c}
\left(\lambda+1+M/r+\sqrt{1-2M/r}\left(\ell+\sqrt{\bar{r}/r}\right)
\right)\left(\ell\sqrt{1-2M/r_p}-M/(r_p+r_p\sqrt{1-2M/r_p})\right)
(\bar{r}_p/\bar{r})^\ell\\
-\left(\lambda+1+M/r+\sqrt{1-2M/r}\left(\sqrt{\bar{r}/r}-\ell-1\right)
\right)\left((\ell+1)\sqrt{1-2M/r_p}+M/(r_p+r_p\sqrt{1-2M/r_p})\right)
(\bar{r}/\bar{r}_p)^{\ell+1}
\end{array}
\right\}\Biggr],\nonumber
\end{eqnarray}
where the upper expressions apply in the case $\bar r>\bar r_p$,
and the lower for $\bar r<\bar r_p$. For radial infall
$\dot r_p=-(1-2M/r_p)\sqrt{2M/r_p+\epsilon_0^2-1}/\epsilon_0$.
Note that the choice of the freezing condition for $\psi_f$ ensures
that $\dot\psi_c$ vanishes on the horizon. This would not be true if
we had taken either the particle limit of the Misner\cite{misner} or
Brill-Lindquist\cite{bl} solutions for $\psi$.
It is also worth stressing that
$\dot\psi_c$ does not depend on $\Gamma_{hom}(r_0)$.

In principle, it could turn out that the CFL prescription and the
convective prescription are the same. In Fig.\ \ref{fig:dIdt} we show that
this is not the case. This figure shows $\dot I_{\rm conf}$ for a
particle dropped from rest at $r_0=15(2M)$.  For the $t=0$
hypersurface, with the particle at $r_p=15(2M)$ (equivalent to
$r_p^*=17.64(2M)$), the solution is momentarily stationary, and hence
$I_{\rm conf}$ is momentarily stationary, that is, $\dot I_{\rm
conf}=0$.  Plots are given of $\dot I_{\rm conf}$ on hypersurfaces a
short time after the initial hypersurface, constant-$t$ hypersurfaces
corresponding to particle positions $r_p^*=14.55(2M)$ and
$r_p^*=11.15(2M)$.  The figure shows clearly that $\dot I_{\rm conf}$
does not vanish.

\subsection{Numerical results}

For some purposes the most important question to ask about a
prescription for late time hypersurface data is whether it leads to a
reasonably accurate estimate of the radiated gravitational wave
energy that is generated in an
astrophysical event. For the problem of a particle falling in from
rest at $r_0$, at time $t=0$, we know with considerable accuracy the
energy radiated in each multipole during the infall. We now must ask:
if we replace the evolved data on a late hypersurface by some
prescription (CFL, convective,\ldots), what radiated energy will we
calculate?  A numerical answer to this question is shown in Fig.\ 
\ref{fig:energy}. For a particle falling from $r_0=15(2M)$, the
``true'' quadrupole gravitational radiation emitted is
$1.64\times10^{-2}m_0^2/(2M)$.  Figure \ref{fig:energy} shows the
radiation that is computed if the evolved data on a hypersurface is
replaced with prescribed data and evolved forward in time. The
prescribed data used for the 3-geometry is conformally
flat with a frozen horizon boundary condition. The value of $\dot\psi$
is given by the convective prescription. The results show that the use
of such prescribed data, even at fairly late times, gives a very good
approximation to the radiated energy. The worst case is an
overestimate of the energy by a factor of around 2.4. This is to be
contrasted with energy computed for prescribed CFL data, shown in
Fig.\,\ref{fig:reven}, where the error grows without bound as the
hypersurface is taken to late times.
The result is significantly in error only for data prescribed on a
hypersurface when the particle radial position $r_p$ is around the
peak of the Zerilli potential in Eq.\ (\ref{zpotential}), where we
would expect the largest deviations in results based on any simple
prescription.  To clarify the relationship of the Zerilli potential
and errors in the computed energy, the Zerilli potential is also
indicated in Fig.\ \ref{fig:energy}.

The success of the new prescription is demonstrated in Figs.\ 
\ref{fig:psidot1.5} and \ref{fig:psidot15}. These figures compare the
new prescription (conformally flat, frozen-horizon, convective) for
hypersurface data with the true (i.e., evolved) data and with the BY
data, on several hypersurfaces.  In
Figs.\,\ref{fig:psidot1.5}--\ref{fig:psidot15} we use BY data as our
example of the CFL prescription, since BY is the most familiar case.
There would be negligible differences in appearance if we used instead
any of the other CFL variations studied in Paper II.

Figures \ref{fig:psidot1.5} shows
$\partial\psi/\partial t$ for a particle falling from rest from
position $r_0=1.5(2M)$ at time $t=0$. This is not a reasonable
astrophysical scenario, but it serves to emphasize strong field
effects and magnify differences in hypersurface data.  The improvement
of the new prescription over BY data is dramatic, even on a
hypersurface ($t=2.2[2M], r_p=1.34[2M])$ soon after the initial
hypersurface. The new prescription agrees very well with the evolved
data, both in its general shape near the particle, and in its
numerical values; by contrast the BY data disagrees sharply with the
evolved data. This disagreement grows at later times, and at
$t=4.5(2M)$ the BY and evolved curves have little similarity. On the
other hand, the new prescription gives a curve that agrees
qualitatively with the true data, and agrees in size within a factor
of two. The same comparison is made in Figs.\ \ref{fig:psidot15} for a
particle falling from rest at $r_0=15(2M)$ at $t=0$. Here we see that
both the new prescription and BY data agree with the evolved data
reasonably well until the particle approaches the strong field region
near the peak of the Zerilli potential. At smaller radii there are
significant quantitative differences between the data of the new
prescription and the evolved data, but these differences are much
milder than the wide divergence of the BY data from the evolved data.

A comparison of outgoing waveforms appears in Figs.\ \ref{fig:wvfrm}.
The solid curves in both ({\em a}) and ({\em b}) show the true
waveform, seen by an observer at $r_{\rm obs}=1000(2M)$, for the
quadrupolar part of $\psi$, for a particle falling from rest, at time
$t=0$ from position $r_p=15(2M)$ (equivalent to $r_p=17.64[2M]$). The
dotted curve in ({\em a}) shows the outgoing waveform that results if
conformally flat, frozen-horizon, convective data is substituted for
true data on the $t=63.9(2M)$ hypersurface. Similarly the dashed curve
in ({\em a}) shows the waveform if the substitution is made at
$t=93.5(2M)$. The three curves show remarkable agreement after the
start of quasinormal ringing.  Only at early times is there a
noticeable difference: the long shallow dip in the true waveform is
reduced or missing from the waveforms generated by the data of the new
prescription.  The dotted curve in ({\em b}) shows a waveform
comparison at a yet later time. Although the new waveform has roughly
the same magnitude of quasinormal ringing as the true waveform, it has
a large phase shift relative to the true curve.  This can be
understood with the idea that quasinormal ringing is generated near
the peak of the curvature potential. For the dotted curve in ({\em
b}), the particle is placed at a smaller radius than the potential
peak, and has no history of having moved inward through the potential
peak. The quasinormal ringing excited by the prescribed data, then, 
must start at the time of the hypersurface on which prescribed data
is defined. For the true curve, the quasinormal ringing starts earlier,
at the time the particle passed going inward. Numerical experiments
confirm with reasonable accuracy that the phase shift depends on
particle position in this manner.

In the work of Abrahams and Cook\cite{abcook}, and Baker and
Li\cite{bakerli}, the ``close limit'' was used to evolve prescribed
data for the head on collision of equal mass holes. The close limit
approximation applies to a hypersurface late enough that a single
horizon surrounds both colliding holes, and only the large radius
fields of the colliding holes are of importance.  We have seen in
Paper II, however, that in the particle limit this close limit does
not work at {\em any} separation of the holes, if BY prescribed data
is used. The method fails at large separation, because the close limit
conditions fail; the method fails at small separation because the BY
data is a bad representation of the true data. In Fig.\ \ref{fig:clap}
we see that the close limit does work at sufficiently small
separations if the new prescription (conformally flat, horizon frozen,
convective) is used.

\section{Conclusions}

We have seen that the convective prescription for initial data for a
black hole coalescence gives a much better description of the way in
which hypersurface data evolves than does the CFL prescription. The
demonstration, however, has been confined to an example of a
coalescence which (i) was a head on collision (ii) used a conformally
flat 3-geometry, and (iii) involved the particle limit of nonrotating
holes.

It will be relatively simple to study another particle-limit example
that is free of the first two of these  limitations. As already
mentioned in Sec.\ IIB, the steady state solution for a particle in
circular orbit around a hole is an example of a solution that is
convective for any time slice. It is straightforward to find this
convective periodic solution in the form of a Fourier series, and this
solution will not be conformally flat.  This opens the possibility for
several interesting comparisons. It will be reasonably straightforward
to start with CFL data for a particle which at one instant has the
correct position and momentum for a circular orbit. Similarly, it is
simple to find initial conditions corresponding to a convective,
initially conformally flat data. Both these solutions could be evolved
forward in time and the anomalous radiation could be found that is
emitted as the field settles into its steady state solution. The
amount of radiation energy in this anomalous initial burst is an
indicator of how sensitive conclusions will be to the details of the
initial data prescribed.  If the energy is significant compared to the
steady state emission during a single orbit, it is a sign that CFL
initial data will give misleading results for black hole coalescence,
at least  for coalescing holes of very different mass.

A considerably more difficult issue is the particle-limit context of
our development. How are these ideas to be applied to coalescence of
holes of comparable mass?
In a very general way one might try to use the idea of a time sequence
of 3-geometries in order to fix the initial extrinsic curvature.
There are many caveats that go with such a proposal, along with the 
question of just how  one implements such a scheme?  One could, for
example, add time dependence to the conformal factor with the
convective description, by making the parameters (location and
momentum of the holes) time dependent. But we do not know at the
outset what the correct dynamical sequence is for strong field
motion. There is a more serious difficulty. Such a scheme would give
a definitive prescription for finding the time derivatives of the
3-geometries, and (with some physically motivated choice of lapse and
shift) would give us the initial extrinsic curvature. This extrinsic
curvature should in general give a result that does not satisfy the
momentum constraint.

A more promising approach might be to try to generalize the condition
$\dot I_{\rm conf}=0$. One might, for example, use the
Cotton-York\cite{cy} tensor $\beta_{ij}$, which vanishes if and only if
the 3-geometry is conformally flat.  By setting the time derivative of
this tensor to zero one would capture, in the nonlinear theory, the
same condition as $\dot I_{\rm conf}=0$ in perturbation theory. 
A scheme like this, however, is based on conformal flatness, and this
raises the question of whether conformal flatness itself should be
retained in an astrophysically motivated specification of initial
data.

Somewhat related to the question of a convective prescription, but
somewhat independent, is the question of the frozen horizon
condition. For representing initial data that evolved from an earlier
astrophysical configuration, the frozen horizon condition is
``obviously'' a correct constraint on data on a late
hypersurface. Whether or not the convective prescription is used in
connection with numerical relativity, the frozen horizon condition
should be used if one wants a representation of an astrophysical
problem.

\begin{acknowledgments}
This work has been partially supported by the National Science
Foundation under grant PHY0507719.
\end{acknowledgments}



\begin{figure}[h]
\caption{  Radiated energy for replacement of evolved data by prescribed data.
  The specific prescribed data used was (CFL) conformally flat,
  longitudinal data with $\psi$ at the horizon taken to be ``frozen''
  to the starting value, and $\dot{\psi}$ ``matched'' to the starting
  form. A particle was started from rest at $r_0=15(2M)$ with momentarily
stationary data, and the solution for $\psi$ was evolved numerically.
On various constant time hypersurfaces, labeled with the $r_p^*$
position of the particle on the hypersurface, the evolved data 
was replaced by the frozen-matched prescribed data, and evolved forward.
The radiated $\ell=2$ energy is plotted as a function of the hypersurface
label $r_p^*$ at which the replacement was made.}\label{fig:reven}
\end{figure}

\begin{figure}[h]
\caption{Comparison of evolved and prescribed $\psi$ for infall from rest 
  at $r_0=15(2M)$ at $t=0$. The comparison is shown for the
  hypersurface at $t=96.9(2M)$, at which time the particle has fallen
  to $r_p=2(2M)$ and has a momentum $P=1.27m_0$. The prescribed $\psi$
  shown is for CF 3-geometry, with a frozen boundary condition at the
  horizon.}\label{fig:revpsi}
\end{figure}

\begin{figure}[h]
\caption{Comparison of evolved and prescribed $\dot{\psi}$ for
infall from rest at $r_0=15(2M)$ at $t=0$. The comparison is shown for the
hypersurface at $t=96.9(2M)$, at which time the particle has fallen
to $r_p=2(2M)$ and has a momentum $P=1.27m_0$. The prescribed $\dot{\psi}$
shown is for CFL extrinsic curvature with the boundary condition at the
horizon ``matched'' to the conditions at $t=0$.
}\label{fig:revpsidot}
\end{figure}

\begin{figure}
\caption{The time derivative of the index of conformality $\dot I_{\rm conf}$
for the CFL prescription applied on constant-$t$ hypersurfaces, for a
particle that starts from rest at $r_p^*=17.64(2M)$
[$r_p=15(2M)$]. The figure shows that the CFL prescription, in
contrast to the convective prescription, implies a nonvanishing $\dot
I_{\rm conf}$.}\label{fig:dIdt}
\end{figure}

\begin{figure}
\caption{
  Radiated quadrupole energy for a particle falling from rest at
  $r_0=15(2M)$. The energy actually radiated (labeled ``True energy''
  is the horizontal line at $E_2=1.64\times10^{-2}m_0^2/(2M)$. The
  dashed line shows energy computed from convective, conformally flat,
  horizon frozen boundary conditions. The prescribed data was used for
  a hypersurface on which the particle was at position $r_p^*$, and
  evolved forward in time. To underscore the role of the Zerilli
  potential, a scaled plot of the potential is shown as a dotted
  curve. This plot of the potential is shifted vertically so that
  aligns with the energy curve as $r^*\rightarrow\infty$.
  }\label{fig:energy}
\end{figure}

\begin{figure}
\caption{Comparison of the evolved $\dot\psi$ with the new prescription
  (conformally flat, frozen-horizon, convective) and with BY data, on
  three successive hypersurfaces, for a particle started from rest at
  $r_0=1.5(2M)$.  Even at very early stages the BY prescription fails.
  By contrast, the new prescription continues to give good predictions
  for most of the trajectory.}
\label{fig:psidot1.5}
\end{figure}

\begin{figure}
\caption{Comparison of the evolved $\dot\psi$ with the new prescription
  (conformally flat, frozen-horizon, convective) and with BY data, on
  three successive hypersurfaces, for a particle started from rest at
  $r_0=15(2M)$.  At very early stages the BY and new prescriptions
  give reasonable accuracy. At somewhat later times BY predictions are
  greatly in error, while the new prescription continues to give good
  predictions for most of the trajectory outside the peak of the
  potential.}\label{fig:psidot15}
\end{figure}

\begin{figure}
\caption{The waveform as seen by an observer located at $1000(2M)$.
  The solid line is the waveform generated by the infall of a particle
  from rest at $r_0=15(2M)$. The other curves are the result of
  replacing the evolved data with new data (conformally flat,
  frozen-horizon, convective) on hypersurfaces at the times indicated.
  If the replacement is made early enough (before the particle reaches
  the maximum of the Zerilli potential) the excitation of quasinormal
  ringing has the correct magnitude and phase, as can be seen in ({\em
    a}). For a replacement at a later time, shown in ({\em b}), there
  is a shift in the quasinormal ringing. 
}
\label{fig:wvfrm}
\end{figure}

\begin{figure}
\caption{The $\ell=2$ radiated energy as predicted by the extension of
  the close limit (Ref.\ \protect\cite{bakerli}) to the particle case.
  Quadrupolar energy calculated from the close limit, applied to the
  new prescription for data on hypersurfaces at various particle
  position $r_p^*$, are compared with the true radiated energy. These
  close limit results are reasonably accurate at sufficiently small
  separation. This should be contrasted with the situation for CFL
  data, given in Fig.\ 26 of Paper II, where disagreement was
  divergent at small separation}
\label{fig:clap}
\end{figure}

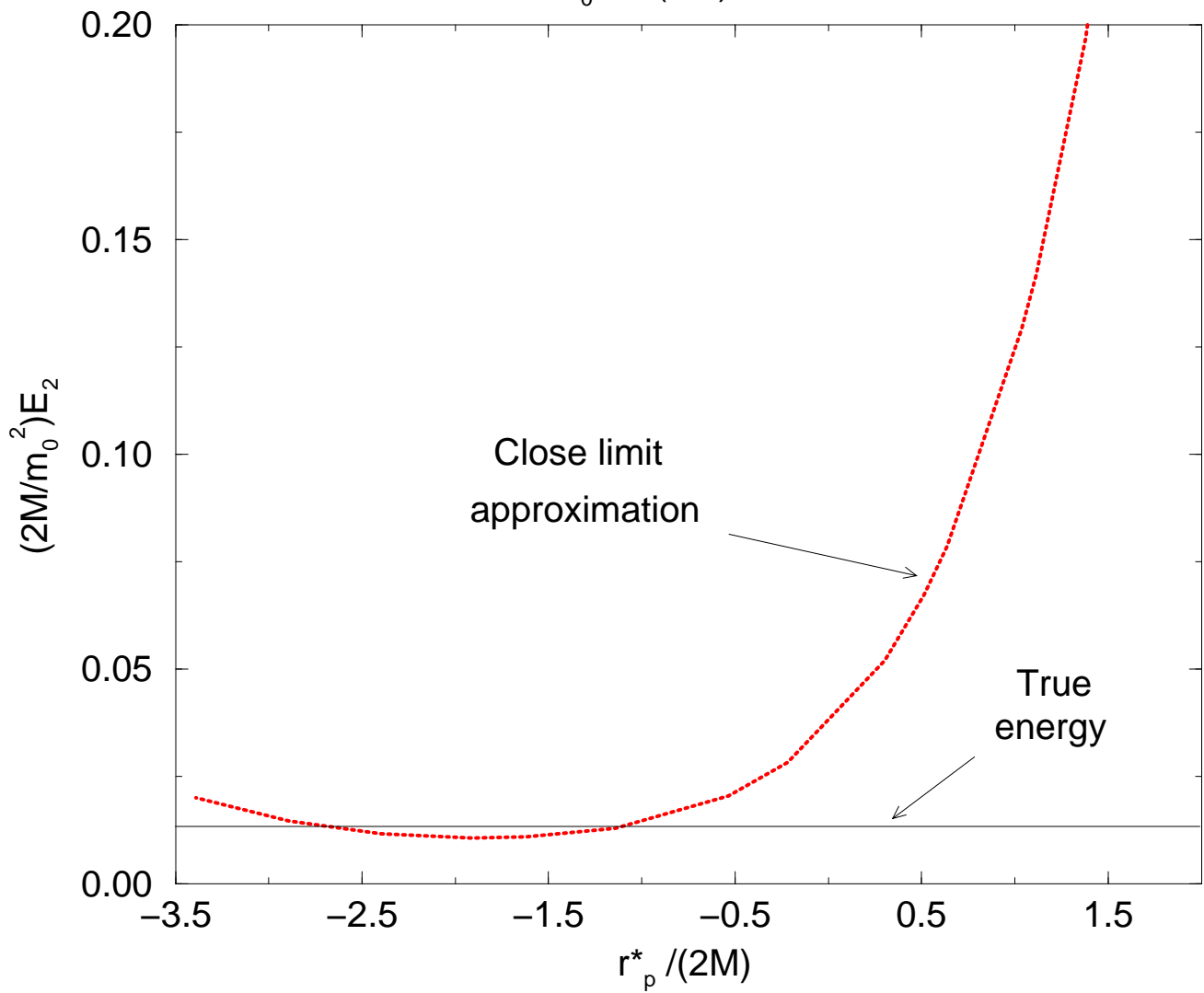
\end{document}